\documentclass[preprint,aps,showpacs,preprintnumbers,amsmath,amssymb]{revtex4}

\usepackage{epsfig}
\usepackage{graphicx}
\usepackage{subfigure}
\usepackage{dcolumn}
\usepackage{bm}
\usepackage{verbatim}

\begin{document}


\title{Shielding of absorbing objects in collisionless flowing plasma}

\author{Yu.~Tyshetskiy}
\email{y.tyshetskiy@physics.usyd.edu.au}
\author{S.~V.~Vladimirov}

\affiliation{%
School of Physics, University of Sydney 2006 NSW Australia
}%

\date{\today}

\begin{abstract}
The electrostatic shielding of a charged absorbing object (dust grain) in a flowing collisionless plasma is investigated by using the linearized kinetic equation for plasma ions with a point-sink term accounting for ion absorption on the object. The effect of absorption on the attractive part of the grain potential is investigated. For subthermal ion flows, the attractive part of the grain potential in the direction perpendicular to the ion flow can be significantly reduced or completely destroyed, depending on the absorption rate. For superthermal ion flows, however, the effect of absorption on the grain attraction in the direction perpendicular to the ion flow is shown to be exponentially weak. It is thus argued that, in the limit of superthermal ion flow, the effect of absorption on the grain shielding potential can be safely ignored for typical grain sizes relevant to complex plasmas.
\end{abstract}

\pacs{52.27.Lw, 52.40.Kh}
\keywords{Dusty plasmas,screening,shielding,plasma kinetic theory}
\maketitle

\section{Introduction}
In the field of complex plasmas~\cite{Vlad_book_05,Fortov_UFN_04,Vladimirov_Ostrikov_04}, one of the central problems is the problem of plasma shielding of charged objects (dust grains) immersed in plasma. One of the effects that influences the shielding is the effect of absorption of plasma particles by a grain, with the latter acting as a sink for plasma particles, creating a flux of electrons and ions from bulk plasma, via the shielding region, to the grain. These fluxes affect plasma far from the grain, preventing it from reaching an equilibrium with the grain field, and hence affecting the grain shielding by the plasma in a qualitative way, compared to a non-absorbing (test) charged grain. This qualitative effect of absorption on the screening is important for interaction of dust grains in dust clouds, clusters and crystals~\cite{Tsytovich_book}, and for related phenomena such as phase transitions in complex plasma structures~\cite{Khrapak_etal_PRL_06,Khrapak_Morfill_PRL_09}.

It was recently shown~\cite{Khrapak_PRL_08} that in an isotropic plasma the shielding of a grain (that is at rest relative to the plasma) is changed qualitatively if absorption of ions by the grain is taken into account: instead of Debye-like potential at large distances from the grain, an $r^{-2}$ tail appears in the potential, in case of collisionless plasmas. However, in reality the plasmas in which dust grains are immersed are usually anisotropic, e.g., due to ion fluxes from bulk plasma to the plasma boundaries (these can be dust-plasma boundaries such as in dust voids~\cite{Goree_etal_PRE_99,VladTsyt_PoP_05}), or due to the motion of dust grains relative to the plasma (we note that the latter situation is equally relevant for satellites moving in the plasma environment of upper ionosphere~\cite{MilVlad_GRL_09}). It is therefore necessary to investigate how absorption of plasma particles by a collecting body (e.g., a dust grain), either immersed in a flowing plasma, or moving in an isotropic plasma, affects the shielding of the body by the plasma.

It is well known~\cite{Cooper_69,Kompaneets_NJP_08} that the potential of a test (non-absorbing) charge, either moving in isotropic plasma or, equivalently, stationary in a flowing plasma, has an attractive part in the direction perpendicular to the axis of the flow; in other words, two test charges of the same sign, placed sufficiently far from each other so that the line connecting them is perpendicular to the plasma flow direction, will attract each other electrostatically. This attraction may play a crucial role in formation and sustaining of plasma dust clusters and crystals~(see \cite{Vlad_book_05,Tsytovich_book,Fortov_UFN_04,Vladimirov_Ostrikov_04} and references therein). The attractive part appears due to the $r^{-3}$ tail in the potential of the test charge far from the charge, which has the sign opposite to that of the test charge. It is reasonable to expect that, as in case of isotropic screening, absorption will lead to a $r^{-2}$ tail in the potential, of the sign opposite to that of the $r^{-3}$ tail of the test charge potential, which will dominate over the $r^{-3}$ tail at large distances from the grain, and hence may either reduce or completely destroy the attractive part of the potential, qualitatively changing the interaction of grains perpendicularly to the flow direction.

The effect of absorption on grain shielding in plasmas with a flow has so far only been studied using a hydrodynamic model in Refs.~\cite{Chaudhuri_etal_07,Filippov_etal_09}. It was demonstrated that, at least in the case of highly collisional plasmas considered there, the far asymptote of the grain field is dominated by the effect of absorption. However, it is yet not known how the absorption affects the grain shielding in plasmas with arbitrary degree of collisionality, in particular, in weakly collisional or collisionless plasmas. To answer this question, a kinetic model of grain shielding, accounting for the absorption, should be used. However, the kinetic model for shielding of grains in anisotropic plasma developed by Ivlev~\textit{et al.}~\cite{Ivlev_etal_05}, does not account for the absorption of ions by the grain. It is thus the aim of the present paper to study, using a kinetic model, the effect of absorption on grain shielding in the limiting case of collisionless flowing plasmas, thus covering the opposite limit of collisionality to that studied in~\cite{Chaudhuri_etal_07,Filippov_etal_09}. In particular, we study here how the absorption affects the attractive part of the grain potential (in the direction perpendicular to the anisotropy axis) in collisionless plasmas. The answer to the latter question is important, e.g., for feasibility of the experiment proposed by Kompaneets~\textit{et al.}~\cite{Kompaneets_NJP_08} aimed at observing the attractive part of the grain potential in the direction perpendicular to the ion flow.

\section{Model and governing equations}
We consider a small spherical absorbing body (grain) of charge $Q_d$, either moving in isotropic homogeneous collisionless plasma with velocity $\mathbf{u}$, or immersed in a homogeneous collisionless plasma with a uniform flow $\mathbf{u}$. Stationary kinetic equation for ions in the reference frame of the grain is
\begin{equation}
\mathbf{v}\cdot\frac{\partial f}{\partial \mathbf{r}} - \frac{e}{m_i}\nabla\phi(\mathbf{r})\cdot\frac{\partial f}{\partial\mathbf{v}} = -\delta(\mathbf{r}) v\sigma(v) f,   \label{eq:general_kinetic}
\end{equation}
where $f=f(\mathbf{r},\mathbf{v})$ is the ion distribution function, $\phi(\mathbf{r})$ is the self-consistent potential of the grain in plasma. The term on the right-hand side (rhs) of~(\ref{eq:general_kinetic}) is the point-sink approximation of ion collection by the grain~\cite{Filippov_etal_09,Khrapak_PRL_08}, in which $\delta(\mathbf{r})$ is the Dirac delta-function, and $\sigma(v)$ is the ion collection cross-section of the body, which we assume to be isotropic, i.e., independent of the direction of ion velocity and only depending on its absolute value, $\sigma(\mathbf{v})=\sigma(v)$. The assumption of isotropic ion collection cross-section is valid as long as the field of the grain, in which the ions being collected by the grain move, is approximately central, i.e., the anisotropy of the grain field in the flowing plasma is small in the region of size $\sim\sqrt{\sigma}$ around the grain in which the trajectories of the ions being collected by the grain are significantly modified by its field. The rough criterion of validity of this approximation can be expressed as
\begin{equation}
\frac{\sqrt{\sigma}}{\lambda_a} < 1,   \label{eq:sigma_criterion_gen}
\end{equation}
where $\lambda_a$ is the characteristic length scale at which the anisotropy in shielding of the grain by plasma becomes significant.

In collisionless plasma $\sigma(v)$ can be obtained from conservation of ion energy and angular momentum in the central field of the grain, and for small grains it is well approximated by the Orbital Motion Limited (OML) theory~\cite{Fortov_PR_05},
\begin{equation}
\sigma(v) = \sigma_{\rm OML}(v) = \pi a^2 \left(1-\frac{2 e \phi_s}{m_i v^2}\right),  \label{eq:sigma_OML}
\end{equation}
where $\phi_s$ is the surface potential of the grain, and $a$ is the grain radius. Note that $\sigma_{\rm OML}(v)$ does not depend on the distribution of potential in the viccinity of the grain. This is another reason for using the isotropic approximation $\sigma(\mathbf{v})=\sigma(v)$ for ion absorption by the grain.

Plasma electrons are assumed to be Boltzmann distributed, $n_e=n_0\exp(e\phi/T_e)$, where $n_0$ is the unperturbed plasma density ($n_{e0}=n_{i0}=n_{0}$), and $T_e$ is the electron temperature in energy units. The electron flux absorbed by the grain is assumed to compensate the absorbed ion flux, so that the grain charge $Q_d$ remains constant. The set is coupled by the Poisson's equation:
\begin{equation}
-\nabla^2\phi = 4\pi e (n_i - n_e) + 4\pi Q_d \delta(\mathbf{r}),  \label{eq:Poisson_general}
\end{equation}
where $n_i = \int{f d\mathbf{v}}$ is the ion number density, and $Q_d \delta(\mathbf{r})$ approximates the charge density of the (small) grain, which is located at $\mathbf{r=0}$. This delta-function approximation is well justified when the grain size is small compared to the effective length scale of plasma screening.

We employ the linear response formalism to find the static potential $\phi_p(\mathbf{r})$ induced in the plasma by the absorbing charged grain. In the absence of the grain, the plasma is assumed homogeneous and quasineutral, with no electric field ($\phi_0=0$), and the ion distribution function $f_0(\mathbf{v})$. The grain perturbs the plasma, inducing an electric field $-\nabla \phi_p(\mathbf{r})$ in the plasma, and perturbing the distribution functions of ions, $f(\mathbf{r,v})=f_0(\mathbf{v}) + f_p(\mathbf{r,v})$. Assuming this perturbation to be small, $|f_p|\ll f_0$, we linearize~Eqs~(\ref{eq:general_kinetic})--(\ref{eq:Poisson_general}), solve them using Fourier transform, and obtain the self-consistent potential of the plasma perturbed by the grain in the form:
\begin{eqnarray}
\phi_p(\mathbf{r}) &=& \frac{Q_d}{2\pi^2}\int{\frac{\exp(i\mathbf{k}\cdot\mathbf{r})}{k^2 D(\mathbf{k})} d\mathbf{k}} + \frac{i e}{2\pi^2}\int{\frac{\exp(i\mathbf{k}\cdot\mathbf{r})}{k^2 D(\mathbf{k})} \delta n_{\rm abs}(\mathbf{k}) d\mathbf{k}} \label{eq:phi(r)_gen} \\
&\equiv & \phi_{Q_d}(\mathbf{r}) + \phi_{\rm abs}(\mathbf{r}), \nonumber
\end{eqnarray}
where $\phi_{Q_d}(\mathbf{r})$ is the potential of the test (non-absorbing) charge $Q_d$, and $\phi_{\rm abs}(\mathbf{r})$ is the additional term due to absorption of ions on the grain, in which
\begin{equation}
\delta n_{\rm abs}(\mathbf{k}) = \int{\frac{v\sigma(v) f_0(\mathbf{v})}{\mathbf{k\cdot v}-i0}d\mathbf{v}}.   \label{eq:dn_abs}
\end{equation}
The dielectric function $D(\mathbf{k})$ in (\ref{eq:phi(r)_gen}) is given by
\begin{equation}
D(\mathbf{k}) = 1 + \frac{k_{De}^2}{k^2} - \frac{\omega_{pi}^2}{n_0 k^2}\int{\frac{\mathbf{k}\cdot\partial f_0(\mathbf{v})/\partial\mathbf{v}}{\mathbf{k\cdot v}-i0}d\mathbf{v}},   \label{eq:D(k)}
\end{equation}
where $\omega_{pi} = \sqrt{4\pi e^2 n_0/m_i}$ is the ion plasma frequency, and $k_{De}=\lambda_{De}^{-1}$ is the inverse electron Debye length, $\lambda_{De}=v_{Te}/\omega_{pe}$. The imaginary part $-i0$ in (\ref{eq:D(k)}) and (\ref{eq:dn_abs}) appears due to causality and specifies the rule for going around the pole when integrating over $\mathbf{v}$. When the unperturbed ion distribution function $f_0(\mathbf{v})$ is approximated with a shifted Maxwellian in the reference frame of the grain,
\begin{equation}
f_0(\mathbf{v}) = n_0 \Phi_M(|\mathbf{v}-\mathbf{u}|) = \frac{n_0}{(2\pi v_{Ti}^2)^{3/2}}\exp\left(-\frac{|\mathbf{v}-\mathbf{u}|^2}{2 v_{Ti}^2}\right), \label{eq:f0}
\end{equation}
where $v_{Ti}$ is the thermal velocity of plasma ions, the dielectric function $D(\mathbf{k})$ becomes:
\begin{equation}
D(\mathbf{k}) = 1 + \frac{k_{De}^2}{k^2} + \frac{k_{Di}^2}{k^2}\left[1 + i\sqrt{\frac{\pi}{2}}\left(\frac{\mathbf{k\cdot u}}{k v_{Ti}}\right) W\left(\frac{\mathbf{k\cdot u}}{\sqrt{2} k v_{Ti}}\right)\right],  \label{eq:D(k)_Maxwell}
\end{equation}
where $W(\zeta) = \exp(-\zeta^2) {\rm erfc}(-i\zeta)$ is the plasma dispersion function of a real argument.

Below we consider analytic solutions for $\phi_p(\mathbf{r})$ of Eq.~(\ref{eq:phi(r)_gen}) in the limits of subthermal ($u/v_{Ti}\ll 1$) and superthermal ($u/v_{Ti}\gg 1$) flow velocities.

\section{Results}
\subsection{Subthermal flows, $u\ll v_{Ti}$}
This limit is relevant to a situation where a grain is suspended in a weakly anisotropic collisionless plasma with a subthermal flow (e.g., a grain in collisionless presheath), or to a grain slowly moving (compared to the plasma ion thermal velocity) in isotropic collisionless plasma.

Introducing $\zeta=\mathbf{k\cdot u}/\sqrt{2}kv_{Ti}$ and using the small argument expansion of $W(\zeta)$, we obtain for the dielectric function (\ref{eq:D(k)_Maxwell}) in the limit of subthermal flows:
\begin{equation}
D(\mathbf{k}) = 1 + \frac{k_{D}^2}{k^2}\left[1 - 2\zeta^2 + i\sqrt{\pi}\zeta\right] + O(\zeta^3),   \label{eq:D(k)_sub}
\end{equation}
where $k_{D}\equiv\sqrt{k_{Di}^2 + k_{De}^2}$ is the inverse total (electron and ion) Debye length. Note that in deriving (\ref{eq:D(k)_sub}) we assumed that electrons are hotter than ions, so that $k_{De}\ll k_{Di}$ and hence $k_{D}\approx k_{Di}$.

To calculate $\delta n_{\rm abs}(\mathbf{k})$ for $u<v_{Ti}$, we use the approximation of the shifted Maxwellian~(\ref{eq:f0}) for small $u/v_{Ti}$:
\begin{equation}
f_0(\mathbf{v}) \approx n_0 \Phi_M(v) \left(1+\frac{u v_\parallel}{v_{Ti}^2}\right).
\end{equation}
With this approximation, we have for $\delta n_{\rm abs}(\mathbf{k})$:
\begin{equation}
\delta n_{\rm abs}(\mathbf{k}) = n_0 \int{\frac{v\sigma(v)}{\mathbf{k\cdot v}-i0}\Phi_M(v)d\mathbf{v}} + \frac{n_0 u}{v_{Ti}^2} \int{\frac{v_\parallel v\sigma(v)}{\mathbf{k\cdot v}-i0}\Phi_M(v)d\mathbf{v}}.  \label{eq:dn_abs_sub_2term}
\end{equation}
Performing integration analytically where possible, we have (see Appendix~\ref{app:dn_abs_sub})
\begin{equation}
\delta n_{\rm abs}(\mathbf{k}) = \frac{2i\sqrt{\pi}n_0}{k}\int_0^\infty{\xi^2 \sigma(\xi) \exp(-\xi^2) d\xi}, \label{eq:dn_abs_int}
\end{equation}
where $\xi=v/\sqrt{2}v_{Ti}$. Finally, using (\ref{eq:sigma_OML}) and integrating over $\xi$, we obtain:
\begin{equation}
\delta n_{\rm abs}(\mathbf{k}) = \frac{i\pi n_0}{2k}\pi a^2 (1+2 z \tau),  \label{eq:dn_abs_sub}
\end{equation}
where $\tau=T_e/T_i$ is the ratio of electron and ion temperatures, and $z = e|\phi_s|/T_e$ is the dimensionless surface potential of the grain in plasma (here $T_e$ is the electron temperature in energy units).

\begin{figure}
\includegraphics[width=6.4in]{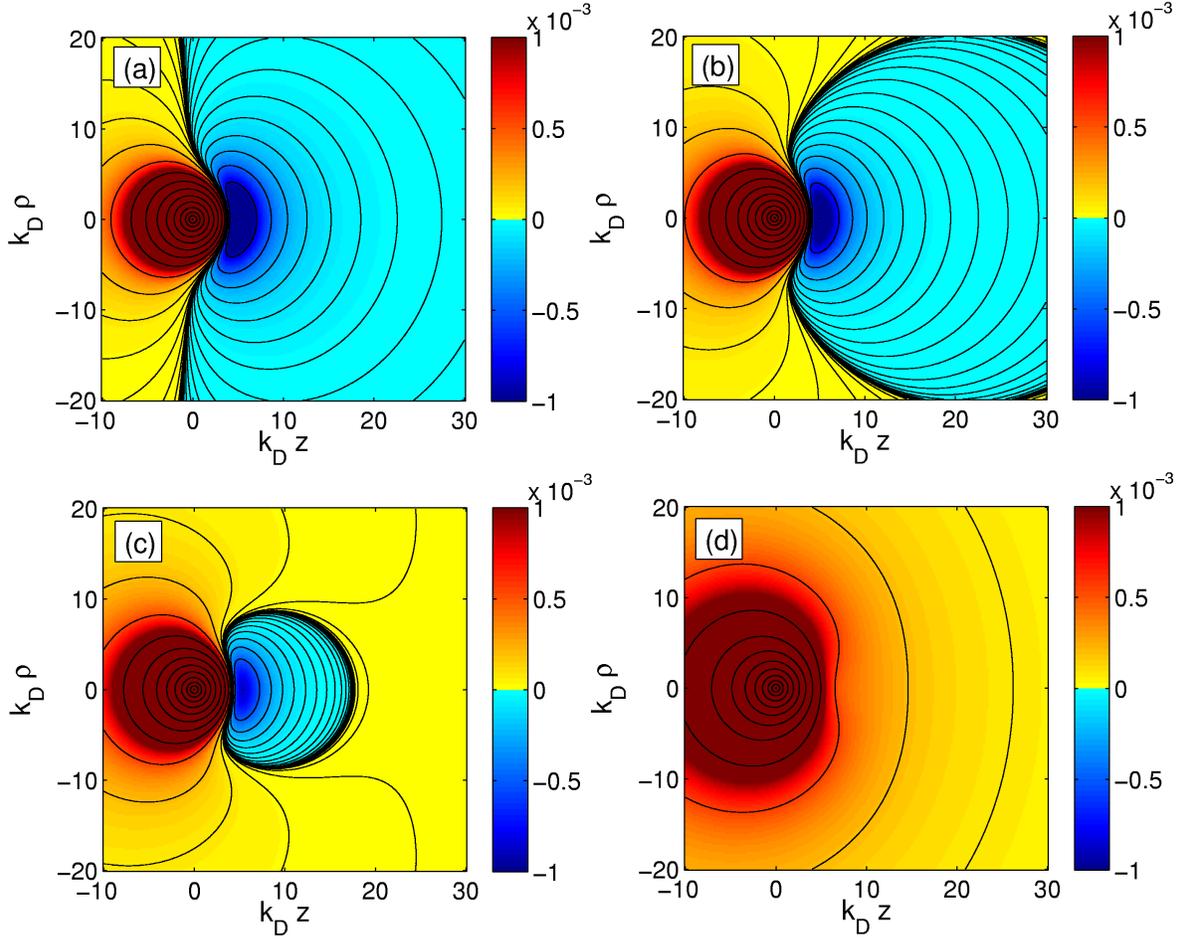}
\caption{\label{fig:phi_Q+a_sub} Grain total potential $\phi_{Q_d}+\phi_{\rm abs}$ (normalized to $Q_d/\lambda_D$), in case of subthermal flow, for different grain sizes (i.e., different rates of absorption): panel (a): $k_Da=0$ (no absorption), panel (b): $k_Da=10^{-2}$, panel (c): $k_Da=2.5\cdot 10^{-2}$, panel (d): $k_Da=10^{-1}$. The grain is located at $\rho=z=0$, the ion flow $\mathbf{u}$ is directed from left to right. The $\phi=0$ contour is shown with a thick black solid line on panels (a)-(c) [on panel (d) the potential does not change sign anywhere]. All potentials are plotted for $u/v_{Ti}=0.2$, $\tau=50$, $z=3$, $n_0=10^8$ cm$^{-3}$.}
\end{figure}

Inserting (\ref{eq:D(k)_sub}) and (\ref{eq:dn_abs_sub}) into (\ref{eq:phi(r)_gen}), expanding up to the second order in $u/v_{Ti}$, and employing spherical coordinates centered at the grain with the axis $\theta=0$ directed along $\mathbf{u}$, we obtain for $\phi_{Q_d}$ and $\phi_{\rm abs}$ for $u<v_{Ti}$ (see Appendix~\ref{app:phi_sub}):
\begin{eqnarray}
\phi_{Q_d}(\mathbf{r}) &\approx & \frac{Q_d}{r}{\rm e}^{-k_Dr} \nonumber \\
&+& \sqrt{\frac{2}{\pi}}Q_d k_D^2 r\cos\theta\left\{-\frac{1}{2k_D^2r^2} + \left(1-\frac{1}{k_Dr}+\frac{1}{k_D^2r^2}\right)\frac{{\rm e}^{k_Dr}}{4k_Dr}E_1(k_Dr)\right. \nonumber \\
&+&\left. \left(1+\frac{1}{k_Dr}+\frac{1}{k_D^2r^2}\right)\frac{{\rm e}^{-k_Dr}}{4k_Dr}{\rm Ei}(k_Dr)\right\}\frac{u}{v_{Ti}} \nonumber\\
&-& \frac{\pi}{2}Q_dk_D\left\{ \frac{1+k_Dr}{24}{\rm e}^{-k_Dr}\right. \nonumber \\
&-&\left.\frac{3\cos^2\theta-1}{k_D^3r^3}\left[1-{\rm e}^{-k_Dr}\left(1+k_Dr+\frac{k_D^2r^2}{2}+\frac{k_D^3r^3}{6}+\frac{k_D^4r^4}{24}\right)\right]\right\}\frac{u^2}{v_{Ti}^2},  \label{eq:phi_Q_sub} \\
\phi_{\rm abs}(\mathbf{r}) &\approx & -\frac{\phi_{a0}}{2k_Dr}\left[{\rm e}^{-k_Dr}\ {\rm Ei}(k_Dr) - {\rm e}^{k_Dr}\ {\rm Ei}(-k_Dr)\right] \nonumber \\
&-& \left(\frac{\pi}{2}\right)^{3/2}\frac{\phi_{a0}\cos\theta}{k_D^2r^2}\left\{1 - {\rm e}^{-k_Dr}\left(1+k_Dr+\frac{k_D^2r^2}{2}\right) \right\} \frac{u}{v_{Ti}} \nonumber\\
&+& \pi\frac{\phi_{a0}k_D^4}{2r}\left\{\cos^2\theta\int_0^\infty{\frac{\sin(kr)dk}{\left(k^2+k_D^2\right)^3}}\right. \nonumber\\
&&\left. + \frac{1-3\cos^2\theta}{r^2}\int_0^\infty{\frac{\sin(kr)-kr\cos(kr)}{k^2\left(k^2+k_D^2\right)^3}dk}\right\}\frac{u^2}{v_{Ti}^2}, \label{eq:phi_abs_sub}
\end{eqnarray}
where $\phi_{a0}=en_0\cdot\pi a^2(1+2z\tau)$, and $E_1$ and ${\rm Ei}$ are exponential integrals. Note that (\ref{eq:phi_Q_sub}) matches the (somewhat corrected) solution of Cooper~\cite{Cooper_69} for the potential of the slowly moving test charge in isotropic plasma. The structure of the field $\phi_{Q_d}$ of the test (non-absorbing) particle has been analyzed in detail in~\cite{Cooper_69}. Here we will only repeat some features of $\phi_{Q_d}$ relevant to this study, along with the corresponding features of the newly obtained $\phi_{\rm abs}$. The spatial structure of the total grain potential $\phi=\phi_{Q_d} + \phi_{\rm abs}$ is shown in Fig.~\ref{fig:phi_Q+a_sub}.

At small distances from the grain, we have for $\phi_{Q_d}$ and $\phi_{\rm abs}$:
\begin{eqnarray}
\phi_{Q_d} & \approx & \frac{Q_d}{r}\exp(-k_D r), \label{eq:phi_Q_sub_near} \\
\phi_{\rm abs} & \approx & \phi_{a0}\left[\ln(k_D r) + \gamma - 1\right],
\end{eqnarray}
where $\gamma\approx 0.5772$ is the Euler gamma. The ratio $\phi_{\rm abs}/\phi_{Q_d}$ tends to zero as $r\rightarrow 0$, i.e., the potential near the grain is dominated by the isotropic Debye potential (\ref{eq:phi_Q_sub_near}), and the role of absorption on the near-grain potential is minor.

Outside the Debye sphere the behaviour of (\ref{eq:phi_Q_sub}) is given by
\begin{eqnarray}
\phi_{Q_d} &=& \frac{Q_d}{r}\left\{ {\rm e}^{-k_Dr}\left(1-\frac{\pi}{16}\frac{u^2}{v_{Ti}^2}k_D^2r^2\cos^2\theta\right) \right.\nonumber\\
&+&\left. \frac{2u}{v_{Ti}}\frac{1}{(k_Dr)^2}\left[\sqrt{\frac{2}{\pi}}\cos\theta + \frac{u}{2v_{Ti}}\left(1-\frac{\pi}{2}\right)\left(1-3\cos^2\theta\right)\right]\right\},\ \ \ k_Dr>1. \label{eq:phi_Q_out}
\end{eqnarray}
At large distances from the grain ($k_Dr\gg 1$), the term proportional to $r^{-3}$ in (\ref{eq:phi_Q_out}) is dominant over the exponentially decaying term, and the asymptote of $\phi_{Q_d}$ at large distances from the grain is
\begin{equation}
\phi_{Q_d} = \frac{Q_d}{k_D^2 r^3}\frac{2u}{v_{Ti}}\left\{\sqrt{\frac{2}{\pi}}\cos\theta + \frac{u}{2 v_{Ti}}\left(1-\frac{\pi}{2}\right)\left(1-3\cos^2\theta\right)\right\},\ \ \ k_Dr\gg 1.  \label{eq:phi_Q_far}
\end{equation}
The far asymptote of $\phi_{\rm abs}$ of (\ref{eq:phi_abs_sub}) is
\begin{equation}
\phi_{\rm abs} = -\frac{\phi_{a0}}{k_D^2r^2}\left\{1 + \sqrt{\frac{\pi}{2}}\frac{u}{2v_{Ti}}\cos\theta + \frac{u^2}{v_{Ti}^2}\left(1-\frac{\pi}{2}\right)\left(1-2\cos^2\theta\right)\right\},\ \ \ k_Dr\gg 1.  \label{eq:phi_abs_far}
\end{equation}
Comparing the radial dependencies of $\phi_{Q_d}$ and $\phi_{\rm abs}$ in (\ref{eq:phi_Q_far}) and (\ref{eq:phi_abs_far}), we see that in case of subthermal flows, the absorption-induced potential $\phi_{\rm abs}$ dominates in the far-field potential of the grain, i.e., the far field of the grain in plasma is defined by absorption of plasma ions on the grain. [This result, obtained kinetically for collisionless plasmas, qualitatively agrees with the findings of Chaudhuri \textit{et al.}~\cite{Chaudhuri_etal_07} for the dominant role of absorption in the far-field of grains immersed in strongly collisional drifting plasmas.] The absorption of ions by the grain changes the radial dependance of the far asymptote of the total grain potential from $\phi\propto r^{-3}$ (of a non-absorbing object) to $\phi\propto r^{-2}$. The characteristic distance $r_{({\rm abs}>Q)}$ from the grain at which this change occurs is defined from (\ref{eq:phi_Q_far}) and (\ref{eq:phi_abs_far}) and depends on $\theta$:
\begin{eqnarray}
k_D r_{({\rm abs}>Q)} &\sim & k_D\frac{|Q_d|}{\phi_{a0}}\frac{u}{v_{Ti}} \sim \frac{2\lambda_D}{a}\frac{u}{v_{Ti}},\ \ \ \ \ \ \ \ \ \ \ \theta=0, \pi, \label{eq:r_Q>abs_par}\\
k_D r_{({\rm abs}>Q)} &\sim & k_D\frac{|Q_d|}{3\phi_{a0}}\left(\frac{u}{v_{Ti}}\right)^2 \sim \frac{2\lambda_D}{3a}\left(\frac{u}{v_{Ti}}\right)^2,\ \ \ \ \theta=\pi/2.  \label{eq:r_Q>abs_perp}
\end{eqnarray}
Here we have used $z=e\phi_s/T_e$ with the surface potential of the grain approximated by $\phi_s\approx Q_d/a$, and the relation $z\tau\gg 1$ typical for dusty plasmas.

The validity of the isotropic approximation for the ion absorption cross-section, as well as the validity of the linear theory expressions for the potentials $\phi_{Q_d}$ and $\phi_{\rm abs}$ in case of subthermal flows, is discussed in Sec.~\ref{sec:validity}.

\begin{figure}
\includegraphics[width=6.4in]{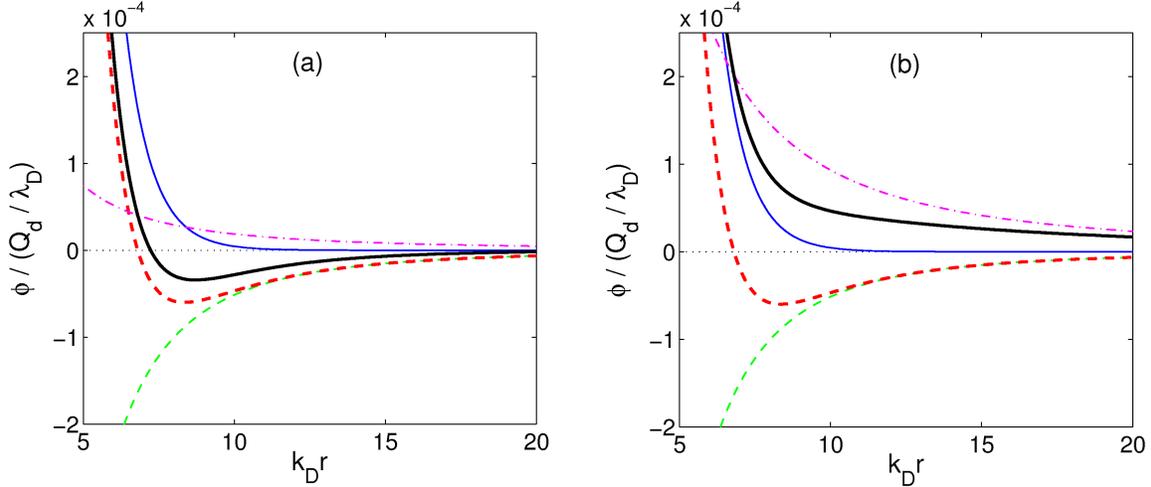}
\caption{\label{fig:phi_Q+a_perp} Potential (normalized to $Q_d/\lambda_D$) in the direction perpendicular to the ion flow, in case of subthermal flow. The thick dashed line shows the potential (\ref{eq:phi_Q_out}) of a non-absorbing (test) charge, consisting of the Debye and $r^{-3}$ components shown with the thin solid and dashed lines, respectively. The thin dashed-dotted line shows the $r^{-2}$ term (\ref{eq:phi_abs_far}) due to absorption of ions by the grain, and the thick solid line shows the total potential $\phi_{Q_d}+\phi_{\rm abs}$ of the absorbing charged grain. The potentials are plotted for $u/v_{Ti}=0.3$, $a/\lambda_D=2\cdot 10^{-3}$ (panel a), and $a/\lambda_D=10^{-2}$ (panel b).}
\end{figure}

\subsection{Superthermal flows, $v_{Ti}\ll u<v_s$}
The limit $v_{Ti}\ll u<v_s$ (where $v_s=(T_e/m_i)^{1/2}$ is the ion sound velocity) can be relevant, e.g., to dust grains suspended against gravity in a collisionless sheath region of the discharge (see, e.g., \cite{Kompaneets_NJP_08,Vladimirov_Ishihara_96,Ish_Vlad_97}), or to satellites in the upper ionosphere~\cite{MilVlad_GRL_09}. In this limit, we can approximate the unperturbed ion distribution function with a shifted delta-function:
\begin{equation}
f_0(\mathbf{v}) = \lim_{v_{Ti}\rightarrow 0}{n_0 \Phi_M(\mathbf{v-u})} = n_0 \delta(\mathbf{v-u}).  \label{eq:f0_super}
\end{equation}
Neglecting the electron response (i.e., assuming that electrons are much hotter than ions so that electrons do not contribute significantly to the screening~\cite{Kompaneets_NJP_08}), the dielectric function (\ref{eq:D(k)_Maxwell}) becomes in this limit:
\begin{equation}
D(\mathbf{k}) = 1 - \frac{1}{(k_{||}-i0)^2 \lambda^2},  \label{eq:D(k)_super}
\end{equation}
where $k_{||}$ is the wavenumber along the direction of $\mathbf{u}$, and $\lambda = u/\omega_{pi} = (u/v_{Ti})\lambda_{Di}$.

Integrating (\ref{eq:dn_abs}) with (\ref{eq:f0_super}), we obtain
\begin{equation}
\delta n_{\rm abs}(\mathbf{k}) = \frac{n_0 u}{\mathbf{k\cdot u}}\sigma(u) = \frac{n_0}{k_{||}}\sigma_{\rm OML}(u),  \label{eq:delta_n_abs_super}
\end{equation}
with $\sigma_{\rm OML}$ defined in (\ref{eq:sigma_OML}).

\begin{figure}
\includegraphics[width=6.4in]{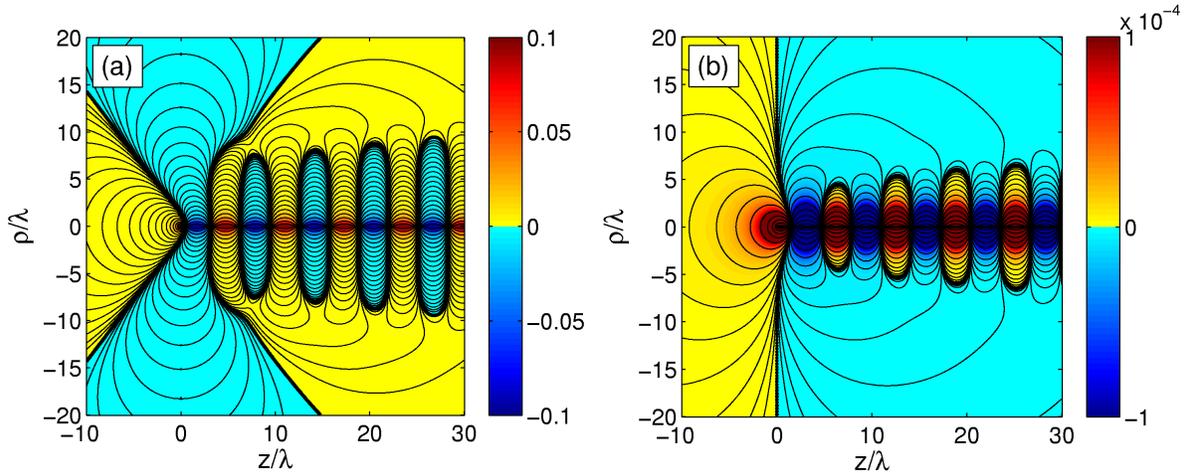}
\caption{\label{fig:phi_super} Plots of $\phi_{Q_d}(\rho,z)$ and $\phi_{\rm abs}(\rho,z)$ given by Eqs~(\ref{eq:phi_Q_super_1})-(\ref{eq:phi_abs_super_2}). Both potentials are normalized by the surface potential of the grain in vacuum, $Q_d/a$. The grain is located at $\rho=z=0$, the flow velocity $\mathbf{u}$ is directed from left to right. The contours represent the equipotential lines (note that here the potential between the contours does not change linearly). The thick black contour shows the line of zero potential, separating the regions of positive and negative $\phi/(Q_d/a)$.}
\end{figure}

Substituting (\ref{eq:D(k)_super}) and (\ref{eq:delta_n_abs_super}) into (\ref{eq:phi(r)_gen}), we obtain for $\phi_{Q_d}$~\cite{Kompaneets_NJP_08} and for $\phi_{\rm abs}$ (see Appendix~\ref{app:phi_super}):
\begin{eqnarray}
\phi_{Q_d}(\rho,z) = \frac{Q_d}{\lambda}\left[\int_{0}^\infty{dt \frac{t^2}{t^2+1} J_0\left(t\frac{\rho}{\lambda}\right) \exp\left(-t\frac{z}{\lambda}\right)} - 2 K_0\left(\frac{\rho}{\lambda}\right)\sin\left(\frac{z}{\lambda}\right)\right],\ \ \ z\geq 0&& \label{eq:phi_Q_super_1} \\
\phi_{Q_d}(\rho,z) = \frac{Q_d}{\lambda}\int_{0}^\infty{dt \frac{t^2}{t^2+1} J_0\left(t\frac{\rho}{\lambda}\right) \exp\left(t\frac{z}{\lambda}\right)},\ \ \ \ \ \ \ \ \ \ \ \ \ \ \ \ \ \ \ \ \ \ \ \ \ \ \ \ \ \ \ \ \ \ \ \ z<0&& \label{eq:phi_Q_super_2}
\end{eqnarray}
and
\begin{eqnarray}
\phi_{\rm abs}(\rho,z) = e n_0 \sigma(u)\left[\int_{0}^\infty{dt \frac{t}{t^2+1} J_0\left(t\frac{\rho}{\lambda}\right) \exp\left(-t\frac{z}{\lambda}\right)} - 2 K_0\left(\frac{\rho}{\lambda}\right)\cos\left(\frac{z}{\lambda}\right)\right],\ \ z\geq 0&& \label{eq:phi_abs_super_1} \\
\phi_{\rm abs}(\rho,z) = -e n_0 \sigma(u)\int_{0}^\infty{dt \frac{t}{t^2+1} J_0\left(t\frac{\rho}{\lambda}\right) \exp\left(t\frac{z}{\lambda}\right)},\ \ \ \ \ \ \ \ \ \ \ \ \ \ \ \ \ \ \ \ \ \ \ \ \ \ \ \ \ \ \ \ z<0&& \label{eq:phi_abs_super_2}
\end{eqnarray}
where $J_0$ is the zero-order cylindrical Bessel function of the first kind, $K_0$ is the zero-order modified Bessel function of the second kind, and $\rho,z$ are the cylindrical coordinates with the grain in the origin $\rho=z=0$ and axis $z$ directed along $\mathbf{u}$. Note that in writing out the Eqs~(\ref{eq:phi_Q_super_1}) and (\ref{eq:phi_abs_super_1}) we omitted the common factor $\exp(-0\cdot z)$ that appears due to the term $-i0$ in (\ref{eq:D(k)_super}), which ensures that the field of the grain vanishes at $z\rightarrow +\infty$ and finite $\rho$. The plots of $\phi_{Q_d}(\rho,z)$ and $\phi_{\rm abs}(\rho,z)$ in case of superthermal flow are shown in Fig.~\ref{fig:phi_super}.

The potential (\ref{eq:phi_Q_super_1})-(\ref{eq:phi_Q_super_2}) of the test charge $Q_d$ was obtained and analysed in detail in~\cite{Kompaneets_NJP_08}. At large distances from the grain, its asymptotic behavior corresponds to a quadrupole potential:
\begin{equation}
\phi_{Q_d} = \frac{Q_d \lambda^2}{r^3}\left(3\cos^2\theta-1\right) + O(r^{-5}),\ \ \ \ \ r\rightarrow\infty,\ \theta\neq 0,  \label{eq:phi_Q_super_as}
\end{equation}
where $\theta$ is the angle between vectors $\mathbf{r}$ and $\mathbf{u}$.

The potential (\ref{eq:phi_abs_super_1})-(\ref{eq:phi_abs_super_2}) due to the effect of particle absorption on the grain has the following asymptotics:
\begin{eqnarray}
\phi_{\rm abs} &=& e n_0 \sigma(u) \frac{\lambda^2\cos\theta}{r^2} + O(r^{-4})\cos^2\theta,\ \ \ \ \text{for }\theta\neq 0, \pi/2, \label{eq:phi_abs_super_as1} \\
\phi_{\rm abs} &=& -e n_0 \sigma(u) \sqrt{\frac{\pi\lambda}{2\rho}}\exp\left(-\frac{\rho}{\lambda}\right),\ \ \ \ \text{for }\theta=\pi/2\ (\text{i.e., }z=0). \label{eq:phi_abs_super_as2}
\end{eqnarray}
As we see, taking into account particle absorption on the grain leads to a qualitative change in the far asymptotic of the grain potential $\phi=\phi_{Q_d}+\phi_{\rm abs}$: instead of the quadrupole potential (\ref{eq:phi_Q_super_as}), it becomes the dipole potential (\ref{eq:phi_abs_super_as1}). The distance $r_{\rm dip}$ at which the dipole potential (\ref{eq:phi_abs_super_as1}) starts to dominate over the quadrupole potential (\ref{eq:phi_Q_super_as}) is
\begin{equation}
\frac{r_{\rm dip}}{\lambda}\sim \left| \frac{Q_d}{e n_0 \sigma(u) \lambda}\right|\frac{3\cos^2\theta-1}{\cos\theta},\ \ \ \theta\neq\frac{\pi}{2},  \label{eq:rdip_lambda}
\end{equation}
Hence in the case of superthermal ion flow the absorption is qualitatively important and should be accounted for, unless $r_{\rm dip}/\lambda\gg 1$. In the latter case, the qualitative change in the grain field asymptote occurs at distances at which the field itself is extremely small and is overwhelmed by other fields. Therefore, in case of superthermal flows, the effect of absorption can be safely ignored for grains for which $r_{\rm dip}/\lambda\gg 1$. After some algebra on Eq~(\ref{eq:rdip_lambda}), the criterion for ignoring the absorption becomes
\begin{eqnarray}
\frac{4z\tau}{1+z\tau v_{Ti}^2/u^2}\frac{\lambda_{Di}^2}{a\lambda}\frac{3\cos^2\theta-1}{\cos\theta} \gg 1, \label{eq:criterion_super}
\end{eqnarray}
where $z=e|\phi_s|/T_e$ is dimensionless grain charge, $\phi_s$ is the surface potential of the grain, $a$ is the grain radius, $\tau=T_e/T_i$ is the ratio of electron and ion temperatures, $\lambda_{Di}$ is the ion Debye length, and $\lambda=u/\omega_{pi}=(u/v_{Ti})\lambda_{Di}$.

The validity of the approximations made and of the obtained potential in the superthermal limit will be discussed in Sec.~\ref{sec:validity}.

\section{Discussion}

\subsection{Modification of attraction between charges of the same sign}
It is easy to see from the asymptotes (\ref{eq:phi_Q_far}) and (\ref{eq:phi_Q_super_as}) of $\phi_{Q_d}$ for sub- and superthermal ion flows, that at $\theta=\pi/2$ the far-field potential $\phi_{Q_d}$ has the sign opposite to the sign of $Q_d$, which implies that a non-absorbing charged grain will attract other grains with charges of the same sign, if they are placed sufficiently far away from the test grain in direction perpendicular to the flow $\mathbf{u}$~\cite{Kompaneets_NJP_08}. Let us analyze how absorption of ions on the grain modifies this attraction, in the limits of subthermal and superthermal ion flows.
\subsubsection{Subthermal ion flows, $u\ll v_{Ti}$ \label{sec:modification_attraction_sub}}
The corresponding attractive potential well of a test (non-absorbing) charge in case of subthermal ion flow is shown by thick dashed lines in Fig.~\ref{fig:phi_Q+a_perp}(a-b). If this potential well is deep enough to confine other dust particles, a crystal-like structure of grains or a grain cluster might form in the plane perpendicular to the ion flow. The distance from the grain to the bottom of the attractive potential well defines the characteristic inter-grain distance in such structure, and the depth of the well defines the \textquotedblleft melting temperature\textquotedblright\ of such structure, i.e., the characteristic kinetic energy necessary for a captured grain to escape the well. [For example, for the potential well shown in Fig.~\ref{fig:phi_Q+a_perp}, the \textquotedblleft melting temperature\textquotedblright\ of the dust grains of the same size as the test grain is of the order $T_d\sim 6 (a/\lambda_D)^2$~eV. For $a/\lambda_D\sim 0.1$ we have $T_d\sim 6\cdot 10^{-2}$~eV, i.e., a dust grain with characteristic kinetic energy below $0.06$~eV will be trapped in the attractive well of radius $a\sim 0.1\lambda_D$ in the plane perpendicular to the ion flow, at distance $\approx 8 \lambda_D$ from the test grain.]

However, as can be seen from (\ref{eq:phi_abs_far}) for $\theta=\pi/2$, absorption of ions by the grain introduces a term which decays as $r^{-2}$, i.e., slower than the asymptote of $\phi_{Q_d}\propto -Q_d/r^3$ responsible for the attraction along the $\theta=\pi/2$ direction, and has the sign opposite to it. In other words, depending on the parameters, absorption can significantly modify the attractive potential well of the grain, either making it much more shallow (and shifting its bottom further away from the grain) if $r_{({\rm abs}>Q)}\gtrsim r_0$, or destroying it completely if $r_{({\rm abs}>Q)}<r_0$ [here $r_{({\rm abs}>Q)}$ is defined by Eq.~(\ref{eq:r_Q>abs_perp}), and  $r_0$ is the radial distance from the grain center to the bottom of its attractive potential well]. This modification is demonstrated in Fig.~\ref{fig:phi_Q+a_perp}. It can be seen from Fig.~\ref{fig:phi_Q+a_perp} that absorption does not destroy the attractive potential well completely only for very small grains, while for larger grains absorption completely destroys the attractive potential well. Indeed, for $u/v_{Ti}=0.3$ used in Fig.~\ref{fig:phi_Q+a_perp}, the bottom of the potential well of the test charge $Q_d$ is at the distance $r_0\sim 8\lambda_D$ from the grain. The absorption will destroy the attractive potential well as long as the $r^{-2}$ tail of $\phi_{\rm abs}$ starts to dominate over the $r^{-3}$ tail of $\phi_{Q_d}$ at distance less than the bottom of the potential well, i.e., if $r_{({\rm abs}>Q)}<r_0$. Using (\ref{eq:r_Q>abs_perp}), we find that this happens for grains larger than $a/\lambda_D\sim 10^{-2}$, for the parameters of Fig.~\ref{fig:phi_Q+a_perp}. For smaller flow velocities, $u/v_{Ti}<0.3$, absorption destroys the attractive potential well for even smaller grains. Typical range of dust grain sizes in dusty plasmas~\cite{NRL} is $a\sim 0.3 - 20\ \mu$m, while typical range of Debye lengths is $\lambda_D\approx\lambda_{Di}\sim 20 - 200\ \mu$m, hence the typical ratio $a/\lambda_D$ is in the range $a/\lambda_D\sim 10^{-3} - 1$, which only marginally includes the extreme value $a/\lambda_{Di}\sim 10^{-2}$ obtained above.
Thus we can conclude that in the limit of subthermal ion flows, the attraction between non-absorbing (test) charged grains in the direction perpendicular to the flow is completely destroyed if absorption of ions by the grains is taken into account, except for very small grains $a/\lambda_D<10^{-2}$.
\subsubsection{Superthermal ion flows, $v_{Ti}\ll u<v_s$ \label{sec:attraction_superthermal}}
In the limit of superthermal flows, the corresponding attractive potential well in the direction $\theta=\pi/2$ from a non-absorbing charged grain has the minimum located at $r\approx 2.2\lambda$, and its depth is $\approx 0.039 Q_d/\lambda$~\cite{Kompaneets_NJP_08}. Absorption of ions by the grain introduces the term (\ref{eq:phi_abs_super_as2}) in the grain potential asymptote. This absorption-induced term is exponentially small and thus never dominates over the quadrupole asymptote~(\ref{eq:phi_Q_super_as}) of $\phi_{Q_d}$. Thus we can conclude that in case of superthermal ion flows the effect of absorption does not destroy the attractive part of the grain potential in the direction perpendicular to the ion flow, for any grain size and ion absorption rate. This important conclusion is favorable for the experiment proposed by Kompaneets~\textit{et al}~\cite{Kompaneets_NJP_08} on detecting the attractive part of the grain potential in direction perpendicular to the flow. This experiment is to be performed with heavy dust particles levitated deep in the collisionless sheath, for which the limit of superthermal ion flow, considered here, is applicable.

\subsection{Significance of the effect of absorption on the grain shielding potential}
Let us now outline the ranges of parameters for which the effect of absorption on the grain shielding is significant and should be accounted for, and for which it can be safely ignored, in the limits of subthermal and superthermal ion flows.
\subsubsection{Subthermal ion flows}
Fig.~\ref{fig:phi_Q+a_sub} illustrates how absorption modifies the grain potential: as the grain size and the corresponding absorption rate increase, the attractive region around the grain shrinks, narrowing down to a finite-size region behind the grain, and eventually disappearing completely for large enough grains. The characteristic distance $r_{({\rm abs}>Q)}$, at which the absorption-induced term $\phi_{\rm abs}$ starts to dominate over the $1/r^3$ asymptote of the test charge potential $\phi_{Q_d}$, depends on $\theta$ and is given by Eqs~(\ref{eq:r_Q>abs_par})-(\ref{eq:r_Q>abs_perp}). The distance $r_{({\rm abs}>Q)}$ is the smallest along the $\theta=\pi/2$ direction. If the minimum distance $r_{({\rm abs}>Q)}$ is not too large, $r_{({\rm abs}>Q)}(\theta=\pi/2)\lesssim r_0$, where $r_0$ is the distance to the bottom of the attractive potential well in the direction $\theta=\pi/2$, then, as discussed in subsection~\ref{sec:modification_attraction_sub} and seen from Fig.~\ref{fig:phi_Q+a_sub}, the effect of absorption is significant for all grain sizes in the typical range $a/\lambda_D\sim 10^{-3} - 1$: for grain sizes in the range $a/\lambda_D\gtrsim 10^{-2}$ the absorption destroys the attractive part of the potential along the $\theta=\pi/2$ direction, and for smaller grains the absorption modifies the attractive potential well along the $\theta=\pi/2$ direction, making it more shallow. Hence, for grains of all sizes relevant to complex plasmas~\cite{NRL} ($a/\lambda_D\sim 10^{-3} - 1$) immersed in a collisionless plasma with subthermal ion flow, one should account for the absorption-induced correction $\phi_{\rm abs}$ to the grain shielding potential, if one is interested in effects associated with the far asymptote of the grain field in the direction perpendicular to the flow (e.g., grain interactions and phase transitions in dust clouds, clusters, and crystals). [We should note, however, that the approximation of the grain as a point charge and point absorber in Eqs~(\ref{eq:general_kinetic}) and (\ref{eq:Poisson_general}) becomes formally invalid for large grains, $a/\lambda_D\gtrsim 1$, in the subthermal limit. In this case the point approximation can be trivially generalized to account for the finite size of the grain. However, such modification of the model will not affect the far field of the grain, and hence is not considered here.]
\subsubsection{Superthermal ion flows}
In the limit of superthermal ion flows, one can safely neglect the effect of absorption on the grain potential if the criterion (\ref{eq:criterion_super}) is satisfied. Typical parameters~\cite{NRL} relevant to the case of superthermal ion flows considered here (e.g., experiments in which dust grains are suspended against gravity in plasma sheath region with superthermal ion flow) are: $z\sim 1$ (see Appendix~\ref{app:estimate_z_super}), $\tau\sim 10 - 100$, and $u/v_{Ti}\sim u_B/v_{Ti}\sim 3 - 10$ (here $u_B$ is the Bohm velocity of ions in the sheath, which is equal to the ion sound velocity $v_s=\sqrt{T_e/m_i}$, and thus $u/v_{Ti}\sim v_s/v_{Ti}=\sqrt{T_e/T_i}=\sqrt{\tau}$). The criterion (\ref{eq:criterion_super}) then becomes, roughly:
\begin{equation}
\frac{r_{\rm dip}}{\lambda}\sim 10 \frac{\lambda_{Di}}{a} \gg 1,   \label{eq:rdip_lambda_super}
\end{equation}
which is well satisfied for typical grain sizes $a/\lambda_{Di}\sim 10^{-3} - 1$.

We therefore conclude that for typical conditions of Earth-based dusty plasma experiments in which dust grains are suspended in the sheath region of the discharge with superthermal ion flow, it is safe to neglect the effect of absorption of ions on the grain when calculating the potential of the grain in plasma, and just to use the expressions (\ref{eq:phi_Q_super_1})-(\ref{eq:phi_Q_super_2}) for the potential of the test (non-absorbing) charge. Only for very large grains, $a/\lambda_{Di}\gtrsim 10$, suspended in a sheath plasma with superthermal ion flow, one should account for the effect of absorption on the grain potential, given by Eqs~(\ref{eq:phi_abs_super_1})-(\ref{eq:phi_abs_super_2}). Yet we would like to stress once again that, as discussed above in Sec.~\ref{sec:attraction_superthermal}, the attractive part of the grain potential in the direction perpendicular to the superthermal ion flow is only slightly affected by the absorption, and can not be destroyed by it, for any grain size (unlike in the case of subthermal flows).

\subsection{Validity of the model \label{sec:validity}}
Finally, let us discuss the validity of the approximations made in deriving the grain shielding potential in both subthermal and superthermal limits, and define the range of parameters for which the obtained results are valid.

\subsubsection{Linear approximation}
The total grain potential $\phi$ is calculated using the linear perturbation analysis, therefore the validity of the obtained expressions for the potential is limited to the region where the nonlinear effects are negligible, i.e., to distances sufficiently far away from the grain, such that the absolute value of the potential energy of ions in the grain field $|e\phi(\mathbf{r})|$ is small compared to their characteristic kinetic energy $\epsilon_i$. In the region near the grain where the potential energy of ions in the field of the grain becomes comparable to their characteristic kinetic energy, $|e\phi(\mathbf{r})|\gtrsim \epsilon_i$, the shielding becomes nonlinear~\cite{B&R_59,AGP_book,Tsytovich_NL_05}, and this region of nonlinear shielding can be considered as a large \textquotedblleft grain\textquotedblright\ of charge $Q_{\rm eff}$ ($Q_{\rm eff}<Q_d$ because of partial grain shielding in the nonlinear shielding region), which is then shielded linearly by the surrounding plasma. In this case the expressions for the grain potential still remain valid outside the region of nonlinear shielding, but with the true grain charge $Q_d$ replaced with the effective charge $Q_{\rm eff}$. However, if the characteristic size of the region of nonlinear shielding is small compared to the characteristic shielding length, then its effect can be neglected, and $Q_{\rm eff}\approx Q_d$.

Let us estimate the characteristic sizes of the nonlinear screening regions around the grain, corresponding to the two independent parts of the total potential of the grain, $\phi_{Q_d}$ and $\phi_{\rm abs}$. At small distances from the grain, the potentials $\phi_{Q_d}$ and $\phi_{\rm abs}$ in both subthermal and superthermal limits become
\begin{eqnarray}
\phi_{Q_d}&\approx&\frac{Q_d}{r},\ \ \ \ \ \ \ \ \ \ \ \ \ \ \ \ \ \ \ \ \ \ \ \ \ \ \ \ \ r\ll \lambda_s, \label{eq:phi_Q_near}\\
\phi_{\rm abs}&\approx&\phi_{a0}\left[{\rm ln}\left(\frac{r}{\lambda_s}\right) + \gamma - 1\right],\ \ \ r\ll \lambda_s, \label{eq:phi_abs_near}
\end{eqnarray}
where $\phi_{a0}=en_0 \sigma({\rm min}\{u,v_{Ti}\})$, $\lambda_s$ is the corresponding screening length [in the subthermal limit we have $\lambda_s=\lambda_D$ and $\phi_{a0}=en_0 \sigma(v_{Ti})$, in the superthermal limit we have $\lambda_s=\lambda=u/\omega_{pi}$ and $\phi_{a0}=en_0 \sigma(u)$], $\gamma\approx 0.5772$ is the Euler gamma. The characteristic sizes of the nonlinear screening regions around the grain, corresponding to $\phi_{Q_d}$ and $\phi_{\rm abs}$, are, respectively, $r_{NL}^{Q} \sim e Q_d/\epsilon_i$ and $r_{NL}^{a} \sim \lambda_s \exp[1-\gamma-\epsilon_i/(e\phi_{a0})]$ (note that the given expression for $r_{NL}^{a}$ is valid for $r_{NL}^{a}/\lambda_s < \exp(1-\gamma)\sim 1.53$).

In the subthermal limit, $\epsilon_i=T_i$ and $\lambda_s=\lambda_D$, and using (\ref{eq:phi_Q_near})-(\ref{eq:phi_abs_near}), we have for $r_{NL}^Q$ and $r_{NL}^a$:
\begin{eqnarray}
\frac{r_{NL}^Q}{\lambda_D} &\sim & \frac{a}{\lambda_D} z\tau, \\
\frac{r_{NL}^a}{\lambda_D} &\sim & \exp\left(1-\gamma-\frac{4}{1+2z\tau}\frac{\lambda_{Di}^2}{a^2}\right).
\end{eqnarray}
For the typical parameters $z\sim 1$, $\tau\sim 10^2$, the distance $r_{NL}^Q$ is small compared to $\lambda_D$ only for very small grains, $a/\lambda_D<10^{-2}$, while for larger grains the region of nonlinear screening is comparable to or even exceeds $\lambda_D$, i.e., the shielding of the test (non-absorbing) charge is linear only for very small grains. However, as noted above, even for larger grains, for which $r_{NL}^Q \gtrsim \lambda_D$ and the shielding is nonlinear for $r\lesssim r_{NL}^Q$, the linear theory still becomes applicable at $r>r_{NL}^Q$, and the expression (\ref{eq:phi_Q_sub}) for the shielding potential of the test charge is still valid for $r>r_{NL}^Q$, with $Q_d$ replaced by the effective charge $Q_{\rm eff}<Q_d$, which is the remaining unscreened charge of the grain at the distance $r_{NL}^Q$ from the grain center. At the same time $r_{NL}^a$ is small compared to $\lambda_D$ for all grains with $a/\lambda_D\lesssim 0.1$, while for $a/\lambda_D\sim 1$ (the largest grains of the typical range of $a/\lambda_D\sim 10^{-3} - 1$) $r_{NL}^a$ becomes comparable with $\lambda_D$, and we expect some nonlinear modification of $\phi_{\rm abs}$. Hence, the expression (\ref{eq:phi_abs_sub}) for the linear potential $\phi_{\rm abs}$ is valid for grains with $a/\lambda_D \lesssim 0.1$, in subthermal limit. Yet even for large grains $a/\lambda_D\sim 1$, when the linear theory ceases to be valid, we do not expect the attractive part of the grain potential, already destroyed by the effect of absorption for much smaller grains $a/\lambda_D<10^{-2}$, to reappear due to the nonlinear modification of $\phi_{\rm abs}$.

In the superthermal limit, using (\ref{eq:phi_Q_near})-(\ref{eq:phi_abs_near}), we have for $r_{NL}^Q$ and $r_{NL}^a$:
\begin{eqnarray}
\frac{r_{NL}^Q}{\lambda} &\sim & \frac{e^2 Z_d}{\lambda\epsilon_i} \sim \frac{a}{\lambda}\frac{T_i}{\epsilon_i} z\tau, \\
\frac{r_{NL}^a}{\lambda} &\sim & \exp\left(1-\gamma -\frac{2\lambda^2}{a^2}\left(1+z\frac{T_e}{\epsilon_i}\right)^{-1}\right).  \label{eq:r_NL_a_super}
\end{eqnarray}
For $\epsilon_i = 2$~eV (in the sheath, to which the superthermal limit is relevant, ions can have even larger energies), $T_i=0.01$~eV, $T_e\sim\epsilon_i$ (in the sheath, $u\sim v_s =\sqrt{T_e/m_i}$, and hence $T_e/\epsilon_i\sim 1$), $n_0=10^8$~cm$^{-3}$, $Z_d\sim 10^4$, and $z\sim 1$ (see Appendix~\ref{app:estimate_z_super}), we have $r_{NL}^Q/\lambda\sim z a/\lambda \sim (a/\lambda_D)(v_{Ti}/u)\ll 1$ for $a/\lambda_D \sim 10^{-3} - 1$, i.e., the linear theory expressions (\ref{eq:phi_Q_super_1})-(\ref{eq:phi_Q_super_2}) for $\phi_{Q_d}$ are valid for all the grain sizes typical for complex plasmas~\cite{NRL}. As for the ratio $r_{NL}^a/\lambda$, it is extremely small for the wide range of parameters, including those typical for the sheath region where the superthermal limit is applicable, due to the exponent in (\ref{eq:r_NL_a_super}). Hence, the linear expressions~(\ref{eq:phi_abs_super_1})-(\ref{eq:phi_abs_super_2}) for $\phi_{\rm abs}$ in the superthermal limit are valid for all of the practical grain sizes.

\subsubsection{Isotropic approximation for ion absorption}
In our model, we also made the approximation of isotropic cross-section of ion absorption on the grain, $\sigma(\mathbf{v})=\sigma(v)$. The validity of this approximation is defined by the criterion~(\ref{eq:sigma_criterion_gen}).

In case of subthermal ion flows, the distance $\lambda_a$ at which the anisotropy of the grain field becomes significant is roughly defined as the distance at which the $1/r^3$ part of (\ref{eq:phi_Q_out}) starts to dominate over the Debye part of (\ref{eq:phi_Q_out}), which is of the order of several $\lambda_D$ for $u/v_{Ti}<1$, and increases with decreasing $u/v_{Ti}$ (for $u/v_{Ti}=0.1$ we have $\lambda_a\approx 5.8 \lambda_D$, and for $u/v_{Ti}\rightarrow 0$ the shielding becomes isotropic, and $\lambda_a\rightarrow\infty$). Taking $\sigma\sim\pi a^2 (1+2z\tau)$, we obtain from (\ref{eq:sigma_criterion_gen}) the following criterion for validity of the isotropic approximation:
\begin{equation}
\frac{a}{\lambda_D}\ < \frac{\lambda_a/\lambda_D}{\sqrt{1+2z\tau}}.
\end{equation}
For $z\sim 1$, $\tau\sim 10^2$, and $\lambda_a\sim 10\lambda_D$, the isotropic approximation $\sigma(\mathbf{v})=\sigma(v)$ is valid for
\begin{equation}
\frac{a}{\lambda_D} < 1,
\end{equation}
i.e., for most of the typical grain sizes $a/\lambda_D\sim 10^{-3} - 1$. Even for the largest of the typical grains, since in the OML approximation $\sigma(v)$ does not depend on the distribution of the grain potential in the vicinity of the grain, we do not expect a significant effect of the grain field anisotropy on the process of ion absorption on the grain, and hence on the derived absorption-induced term $\phi_{\rm abs}$ in the grain potential.

In case of superthermal ion flows, we use (\ref{eq:sigma_criterion_gen}) with $\lambda_a\gtrsim\lambda$, $\sigma=\sigma(u)$, which gives us the following criterion of validity of the isotropic approximation for ion absorption on the grain:
\begin{equation}
\frac{a}{\lambda_{D}} < \frac{u/v_{Ti}}{\sqrt{1+z T_e/\epsilon_i}}.
\end{equation}
For $z\sim 1$, $u/v_{Ti}\sim\sqrt{\tau}\gg 1$, and $T_e/\epsilon_i\sim 1$, this condition is well satisfied (and hence the isotropic approximation is valid) for all grain sizes from the typical range $a/\lambda_{D}\sim a/\lambda_{Di}\sim 10^{-3} - 1$~\cite{NRL}.

We therefore conclude that the approximation of isotropic cross-section of ion absorption on the grain is well justified for typical grain sizes relevant to complex plasmas.

\subsection{Effect of ion-neutral collisions}
The model used in this paper is collisionless, but it can in principle be generalized to include ion-neutral collisions, which are typically the most frequent collisions in complex plasmas~\cite{Fortov_PR_05}. The effect of ion-neutral collisions will most probably modify the results obtained here, as the collisions lead to an $r^{-2}$ tail in the shielding potential of a test (non-absorbing) grain in drifting plasmas~\cite{Stenflo_etal_73,Kompaneets_etal_pop_07}, and to an $r^{-1}$ tail in the shielding potential of an absorbing grain in isotropic plasmas~\cite{Filippov_etal_07,Khrapak_PRL_08,Khrapak_etal_PRL_07} or drifting strongly collisional plasmas~\cite{Chaudhuri_etal_07}. The generalization of the kinetic model with absorption used in this work, to include collisions, and the study of the effect of absorption on shielding of dust grains in flowing plasmas with arbitrary degree of ion-neutral collisionality, are left for future work.

\section{Conclusions}
In this work we obtained the shielding potential of an absorbing charged grain, which is either moving in a uniform collisionless plasma, or is immersed in a collisionless plasma with a uniform flow, employing linear perturbation analysis of the kinetic equation for ions with point-sink term accounting for the absorption of ions by the grain. The total potential of the grain in plasma is shown to consist of two independent terms: the term $\phi_{Q_d}$ due to the grain charge, and the term $\phi_{\rm abs}$ due to absorption of plasma ions by the grain. Having obtained the general expression for the absorption-induced term $\phi_{\rm abs}$, we analyzed analytically its effect on the grain shielding in the limits of sub- and superthermal ion flows. It has been shown that, in general, absorption leads to a qualitative change in the far field of the grain from a quadupole-like field of a non-absorbing (test) charge to a dipole-like field of the absorbing grain. The characteristic distance from the grain at which this change occurs is obtained in the limits of subthermal and superthermal ion flow in terms of the plasma parameters and the grain size. This distance varies depending on the direction relative to the direction of the flow, and so does the effect of absorption-induced correction $\phi_{\rm abs}$ to the grain shielding potential. The most important absorption-induced effect on the grain shielding is the effect on the attractive part of the grain potential in the direction perpendicular to the flow, that has been previously demonstrated to appear for a test (non-absorbing) charged grain~\cite{Cooper_69,Kompaneets_NJP_08}, and which is important for the grain interaction in structures (e.g., dust clusters and crystals) in complex plasmas.

In the limit of subthermal flows, the absorption of ions by the grain is shown to significantly modify this attractive part of the grain potential.
The character and magnitude of this modification depends on the plasma parameters and the grain size. The quantitative criterion is derived which defines, for given plasma parameters and grain sizes, how the absorption modifies the attractive part of the grain potential: whether it makes the attractive potential well more shallow, thus making the attraction weaker and reducing the associated \textquotedblleft melting temperature\textquotedblright\ of the dust structure in the plane orthogonal to the flow direction, or whether it destroys the attractive part of the grain completely. Our estimates show that in the subthermal limit $u/v_{Ti}\ll 1$ the effect of absorption on the attractive part of the grain potential is significant for all of the typical grain sizes $a/\lambda_D\sim 10^{-3}-1$ relevant to complex plasmas. In particular, we found that in case of subthermal flow the absorption significantly reduces the attractive part of the grain potential for the smallest of the typical grains, $a/\lambda_D \sim 10^{-3}$, while completely destroying it for grains larger than $a/\lambda_D \sim 10^{-2}$. One should thus account for the effect of ion absorption on the grain in plasmas with subthermal ion flow, if one is interested in the effects associated with the asymptotic behavior of the grain potential in the direction perpendicular to the flow, e.g., interaction of grains and phase transitions in dust crystals and clusters in the plane orthogonal to the ion flow.

In the limit of superthermal flows, however, the situation is different from that of the limit of subthermal flows: the absorption can only slightly modify the attractive part of the grain potential in the direction perpendicular to the ion flow, but can not destroy it, for any grain size. Moreover, for typical parameters and grain sizes relevant to complex plasma experiments where dust grains are levitated in the sheath region, the minimum distance at which the absorption-induced change from the quadrupole to the dipole field occurs is so large compared to the characteristic screening length $\lambda$ [see Eq.~(\ref{eq:rdip_lambda_super})], that the grain field at this distance is practically shielded completely, or at least is shielded to the level where it is overwhelmed by random plasma noise fields. Therefore, in case of superthermal ion flows, the effect of absorption on the grain shielding potential can be safely ignored for typical grain sizes $a/\lambda_{Di}\sim 10^{-3} - 1$ occuring in complex plasmas.

\begin{acknowledgments}
The authors thank R.~Kompaneets for useful discussions. This work was supported by the Australian Research Council.
\end{acknowledgments}

\appendix
\section{Derivation of Eq~(\ref{eq:dn_abs_int}) \label{app:dn_abs_sub}}
Using $d\mathbf{v}=v_\perp dv_\perp dv_\parallel d\alpha$, the integral in the second term in the right-hand side of (\ref{eq:dn_abs_sub_2term}) can be written as
\begin{equation}
\int{\frac{v_\parallel v\sigma(v)}{\mathbf{k\cdot v}-i0}\Phi_M(v)d\mathbf{v}} = 2\pi \int_{-\infty}^\infty{dv_\parallel v_\parallel G(v_\parallel^2)\Phi_M(v_\parallel)},
\end{equation}
where
\begin{equation}
G(v_\parallel^2) = \int_0^\infty{dv_\perp v_\perp \frac{v\sigma(v)}{\sqrt{k_\parallel^2v_\parallel^2 - k_\perp^2v_\perp^2}}\Phi_M(v_\perp)},
\end{equation}
with $v=\sqrt{v_\parallel^2 + v_\perp^2}$. Since $G$ is an even function of $v_\parallel$, the above integral is zero, and the second term in the right-hand side of (\ref{eq:dn_abs_sub_2term}) vanishes.

The integral in the remaining term in (\ref{eq:dn_abs_sub_2term}) is
\begin{equation}
\int{\frac{v\sigma(v)}{\mathbf{k\cdot v}-i0}\Phi_M(v)d\mathbf{v}} = \lim_{\nu\rightarrow 0}{i \int{\frac{v\sigma(v)}{\nu + i\mathbf{k\cdot v}}\Phi_M(v)d\mathbf{v}}}. \label{eq:app:lim}
\end{equation}
Using spherical coordinates $d\mathbf{v}=v^2\sin(\theta)dvd\theta d\varphi$ and integrating over $\theta$ and $\varphi$, we reduce the last integral to
\begin{equation}
\int{\frac{v\sigma(v)}{\nu + i\mathbf{k\cdot v}}\Phi_M(v)d\mathbf{v}} = \frac{4}{\sqrt{\pi}k}\int_0^\infty{\xi^2\sigma(\xi)\arctan\left(\frac{\xi}{\Theta}\right)\exp(-\xi^2) d\xi},
\end{equation}
where $\xi=v/\sqrt{2}v_{Ti}$, $\Theta=\nu/\sqrt{2}kv_{Ti}$. Taking the limit $\nu\rightarrow 0$ in (\ref{eq:app:lim}), we obtain the Eq.~(\ref{eq:dn_abs_int}).

\section{On derivation of Eqs~(\ref{eq:phi_Q_sub}) and (\ref{eq:phi_abs_sub}) \label{app:phi_sub}}
To evaluate the integrals on $\mathbf{k}$ appearing after substituting (\ref{eq:D(k)_sub}) and (\ref{eq:dn_abs_sub}) into (\ref{eq:phi(r)_gen}), we expand $\exp(i\mathbf{k\cdot r})$ following Cooper~\cite{Cooper_69}:
\begin{equation}
\exp(i\mathbf{k\cdot r}) = 4\pi\sum_{l=0}^\infty \sum_{|m|\leq l}{i^l j_l(kr) Y_{lm}^*(\theta_k,\varphi_k) Y_{lm}(\theta,\varphi)},  \label{eq:app_exp(ikr)}
\end{equation}
where $j_l$ are the spherical Bessel functions, $Y_{lm}$ are spherical harmonics, $\theta,\varphi$ and $\theta_k,\varphi_k$ are the spherical angular coordinates of the vectors $\mathbf{r}$ and $\mathbf{k}$, respectively, and the asterix denotes complex conjugation. Then, substituting (\ref{eq:app_exp(ikr)}) into (\ref{eq:phi(r)_gen}) and using $d\mathbf{k} = k^2\sin\theta_k dk d\theta_k d\varphi_k$, we perform the integration over $\theta_k$ and $\varphi_k$ and then, where possible, over $k$, obtaining the Eqs~(\ref{eq:phi_Q_sub}) and (\ref{eq:phi_abs_sub}).

\section{Derivation of Eqs~(\ref{eq:phi_Q_super_1})-(\ref{eq:phi_abs_super_2}) \label{app:phi_super}}
Upon substituting (\ref{eq:D(k)_super}) and (\ref{eq:delta_n_abs_super}) into (\ref{eq:phi(r)_gen}), we obtain the following expressions for $\phi_{Q_d}$ and $\phi_{\rm abs}$ in cylindrical coordinates $\rho,z$ (note that $\mathbf{k\cdot r} = k_{\perp}\rho\cos\varphi_k + k_{\parallel} z$, $d\mathbf{k} = k_{\perp}dk_{\perp}dk_{\parallel}d\varphi_k$):
\begin{eqnarray}
\phi_{Q_d}(\rho,z) &=& \frac{Q_d\lambda^2}{\pi}\int_0^\infty{dk_{\perp}k_{\perp}J_0(k_{\perp}\rho)}\int_{-\infty}^\infty{dk_{\parallel}\frac{k_{\parallel}^2\exp(ik_{\parallel}z)}{\left(k_{\parallel}^2 + k_{\perp}^2\right)\left((k_{\parallel}-i0)^2\lambda^2 - 1\right)}}, \label{eq:app:phi_Q_int} \\
\phi_{\rm abs}(\rho,z) &=& \frac{i e n_0}{\pi} \lambda^2\sigma(u)\int_0^\infty{dk_{\perp}k_{\perp}J_0(k_{\perp}\rho)}\int_{-\infty}^\infty{dk_{\parallel}\frac{k_{\parallel}\exp(ik_{\parallel}z)}{\left(k_{\parallel}^2 + k_{\perp}^2\right)\left((k_{\parallel}-i0)^2\lambda^2 - 1\right)}},  \label{eq:app:phi_abs_int}
\end{eqnarray}
where the integration over the azimuthal angle $\varphi_k$ has been carried out using the formula
\begin{equation}
\int_0^{2\pi}{\exp(i k_\perp \rho \cos\varphi_k) d\varphi_k} = 2\pi J_0(k_\perp\rho).
\end{equation}
The integration over $k_\parallel$ in (\ref{eq:app:phi_Q_int}) and (\ref{eq:app:phi_abs_int}) can be performed analytically using the residue theorem and the Jordan lemma. The poles of the integrand are $k_\parallel = \pm i k_\perp$ and $k_\parallel = \pm \lambda^{-1} + i0$, and the contour of integration is closed in the upper (lower) part of the complex plane of $k_\parallel$ for $z\geq 0$ ($z<0$). Performing the integration over $k_\parallel$ and using the formula
\begin{equation}
\int_0^\infty{dk_\perp \frac{k_\perp J_0(k_\perp\rho)}{k_\perp^2+\lambda^{-2}}} = K_0\left(\frac{\rho}{\lambda}\right),
\end{equation}
we obtain Eqs~(\ref{eq:phi_Q_super_1})-(\ref{eq:phi_abs_super_2}).

\section{Estimate of the grain charge in plasma with superthermal ion flow, $u\gg v_{Ti}$ \label{app:estimate_z_super}}
The grain charge reaches equilibrium when the electron and ion currents on the grain compensate each other, $I_e=I_i$. In the superthermal limit, we approximate the ion distribution function with $f_i(\mathbf{v})=n_0\delta(\mathbf{v-u})$, thus we have for the ion current on the grain:
\begin{equation}
I_i = \int{v \sigma(v) f_i(\mathbf{v}) d\mathbf{v}} = n_0 u \sigma(u).
\end{equation}
Using the OML approximation (\ref{eq:sigma_OML}) for $\sigma$, we have
\begin{equation}
I_i = n_0 u \pi a^2\left(1+\frac{Z_d}{2\pi n_0 a \lambda^2}\right).
\end{equation}

For Maxwell-Boltzmann distribution of electrons in the grain field, the electron current on the grain $I_e$ is~\cite{Fortov_PR_05}
\begin{equation}
I_e = \sqrt{8\pi} a^2 n_0 v_{Te} \exp\left(\frac{e\phi_s}{T_e}\right),
\end{equation}
where $\phi_s$ is the surface potential of the grain.

Now, introducing the dimensionless quantity $z=e|\phi_s|/T_e$ (normalized grain surface potential), which is approximately equal to the normalized grain charge, $z\approx e^2 Z_d/aT_e$ (here $Z_d$ is the charge number of the grain), and requiring that $I_e=I_i$, we obtain the equation for $z$:
\begin{equation}
\sqrt{8\pi}\exp(-z) = \frac{\pi u}{v_{Ti}}\sqrt{\frac{\mu}{\tau}}\left(1 + 2 z\tau\frac{v_{Ti}^2}{u^2}\right),  \label{eq:app_eq_for_z}
\end{equation}
where $\mu=m_e/m_i$ is the ratio of electron and ion masses, and $\tau=T_e/T_i$ is the ratio of electron and ion temperatures. The solution $z$ of (\ref{eq:app_eq_for_z}) weakly (logarithmically) depends on the plasma and flow parameters $\mu$, $\tau$, and $u/v_{Ti}$, and is of the order of unity, $z\sim 1$, for typical values of these parameters. For example, for a grain suspended deep in the sheath region we have $u\sim v_s$, where $v_s=\sqrt{T_e/m_i}$ is the ion sound velocity, and hence $u/v_{Ti}\sim\sqrt{\tau}$. Then the equation (\ref{eq:app_eq_for_z}) reduces to
\begin{equation}
\exp(-z) \sim \sqrt{\frac{\pi\mu}{8}}\left(1 + 2 z\right).
\end{equation}
For hydrogen plasma ($\mu\approx 5.44\cdot 10^{-4}$) its solution yields $z\approx 2.45$, while for argon plasma ($\mu\approx 1.36\cdot 10^{-5}$) its solution yields $z\approx 3.90$.


\end{document}